\documentclass[aps,prl,showpacs,twocolumn,reprint,floats,epsfig,pdflatex,superscriptaddress]{revtex4-1}
\usepackage{epsfig}
\usepackage{amsmath}
\usepackage{amsfonts}
\usepackage{graphicx}
\usepackage{amssymb}
\usepackage{amsbsy}
\usepackage{subfigure}
\usepackage{tabularx}

\newcommand{\ket}[1]{|{#1}\rangle}

\begin{document}

\title {Optical Barium Ion Qubit}

\author{Dahyun~Yum} 
\author{Debashis~De~Munshi} 
\author{Tarun~Dutta}

\affiliation{Centre for Quantum Technologies, National
University Singapore, Singapore 117543}
\author{Manas~Mukherjee}
\affiliation{Centre for Quantum Technologies, National University Singapore, Singapore 117543}
\affiliation{Department of Physics, National University Singapore, Singapore 117551. }

\date{\today}

\begin{abstract}

We demonstrate an optical single qubit based on 6$S_{1/2}$ to 5$D_{5/2}$ quadrupole transition of a single Ba$^{+}$ ion operated by diode based lasers only. The resonance wavelength of the 6$S_{1/2}$ to 5$D_{5/2}$ quadrupole transition is about $1762$~nm which suitably falls close to the U-band of the telecommunication wavelength. Thus this qubit is a naturally attractive choice towards implementation of quantum repeater or quantum networks using existing telecommunication networks. We observe continuous bit-flip oscillations at a rate of about $250$ kHz which is fast enough for the qubit operation as compared to the measured coherence time of over $3$~ms. We also present a technique to quantify the bit-flip error in each qubit NOT gate operation.

\end{abstract}


\maketitle
Ion trap based quantum processors are the forerunners in developing new quantum technologies namely, quantum simulators, quantum computers, clocks {\it etc.} \cite{Cirac1995,Roos2012,Nori2009,chou2010}. The basic architecture of a quantum processor comprises of qubits and certain universal quantum gates.  The most rudimentary processor is a single qubit and a single qubit gate which is any rotation in the Bloch sphere. Even though this processor is of hardly any importance for real application, it forms the fundamental building block of a quantum processor. Therefore over the years considerable efforts have been made systematically improve the quality of a single qubit gate \cite{Preskill,Wineland2011,Ryan2009,Olmschenk,Benhelm,Ospel}. The figures-of-merit of a single qubit gate are its addressibility, speed and fidelity. Among the most advance qubits only barium ion optical qubit provides a qubit transition close to the existing optical telecommunication domain of the U-band. This has a potential advantage of using a barium ion qubit readily with existing telecommunication network. However, so far the optical qubit in barium ion could only reach Rabi frequencies of $50~$kHz limited mainly by the laser linewidth and power\cite{Blinov2010}. At this high speed, the fidelity achieved are also limited by technical noise of the laser which so far are only fiber based lasers. 

In this article, all diode based single qubit system for a barium ion optical qubit operating at $1760~$nm wavelength is introduced. As the optical qubit transition is driven by a diode laser, all existing technology typically used for quantum optics can be directly incorporated and if required miniaturized within the same platform.
The article is divided into three sections: starting with a brief description of the experimental setup followed by a vivid analysis of our results related to the single qubit operations and concludes with the achieved figures-of-merit of the system.

\section{Experimental setup}

\begin{figure}
\includegraphics[width=7.5cm]{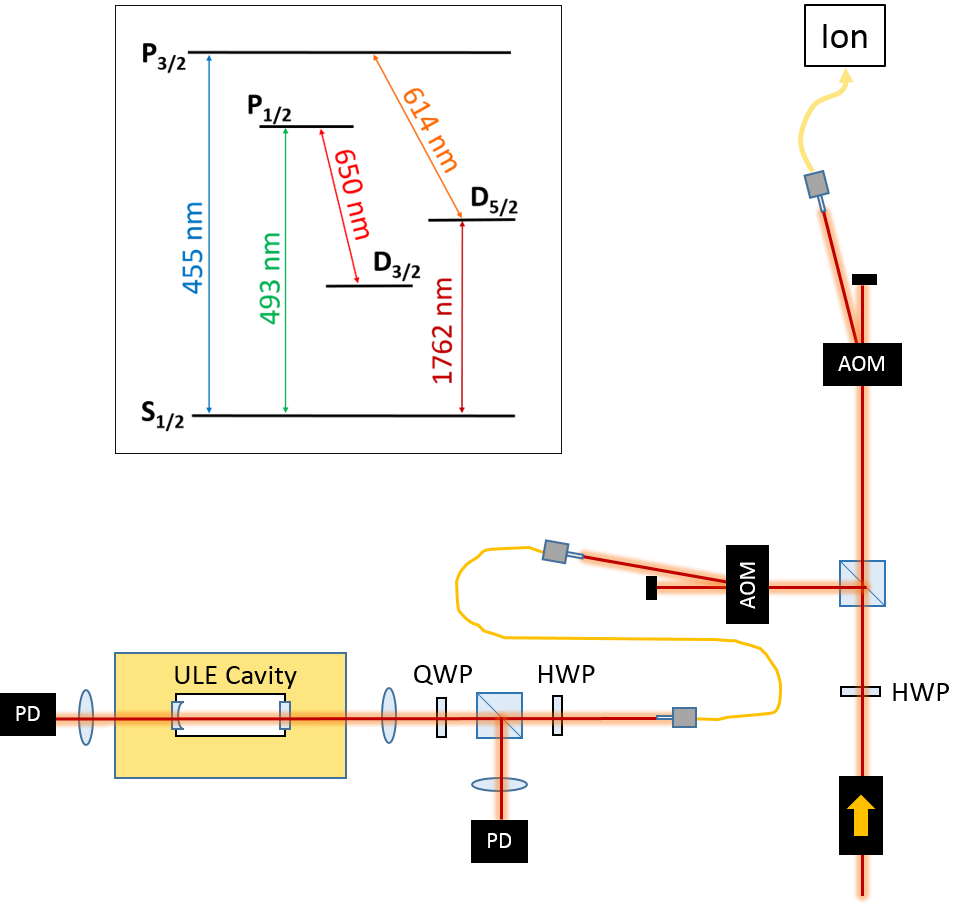}
\caption{(Color Online)Schematic diagram of the $1762~$nm diode laser setup. The laser is phase locked to an ultra low expansion (ULE) cavity to obtain a narrow linewidth stable laser output. An acusto-optic modulator(AOM) is used to controll the laser power and to scan the laser frequency. (inset : Relevant energy level structure of the Ba$^+$ ion.) }\label{fig1}
\end{figure}

At the heart of the experiment is a linear radio-frequency (rf) ion trap with four blade like electrodes for easy optical access. The trap is operated at $16.7~$MHz frequency with the secular motional frequencies of $1.8~$MHz in radial direction while $1~$MHz in the axial direction. A detailed description of the experimental setup can be found in \cite{Manas2015,Manas2016}. The relevant energy levels of a barium ion as shown in Fig.\ref{fig1} consists of a dipole transition at $493~$nm used for Doppler cooling along with a re-pumper at $650~$nm. The qubit in our experiment is the 6S$_{1/2}$-5D$_{5/2}$ optical quadrupole transition which ensures long life of the qubit as the natural decay time of this D-level is about $30~$s \cite{Blinov2014}. The odd isotopes of barium also provides long-lived hyperfine qubits\cite{Blinov2010}, however we focus on the optical qubit only due to its applicability in telecommunication. In order to achieve scalability all the lasers are extended cavity diode lasers (ECDL). In case of the $493~$nm laser, frequency doubling from a $986~$nm ECDL is performed in a cavity while for the $614~$nm re-pumping light, the doubling is performed in a periodically polled optical wave guide from a master laser at $1228~$nm, also in an ECDL format. In order to avoid any drift in the ECDL laser frequencies, the cooling laser is locked to a reference transition line of a tellurium molecule\cite{Tel} while the re-pumper is monitored via barium hollow cathode lamp opto-galvanometric spectroscopy\cite{tarun}. 

The qubit transition is driven by an ECDL at $1760~$nm. The total output power of the laser is about $50~$mW. The power output after a $60~$dB optical isolator is distributed as shown in Fig.\ref{fig1} using commercial telecommunication optical fiber with efficiency as high as $95\%$. A part of the light, about a few hundred $\mu$W, is used for frequency stabilization while another small fraction is fiber coupled to a {\it Bristol Instrument} wavemeter with a resolution of $20~$MHz which serves as a course reference for the atomic transition. An acusto-optic modulator~(AOM) is used to scan the laser frequency before it is transported to the ion trap via a five meter long single mode optical fiber. At the end of the fiber and before the vacuum window of the ion trap, the maximum optical power is over $10~$mW. This power is varied by varying the AOM input rf power which is generated by a direct digital synthesizer (DDS) board via a python based control software. 

\begin{figure}
\includegraphics[width=7.5cm]{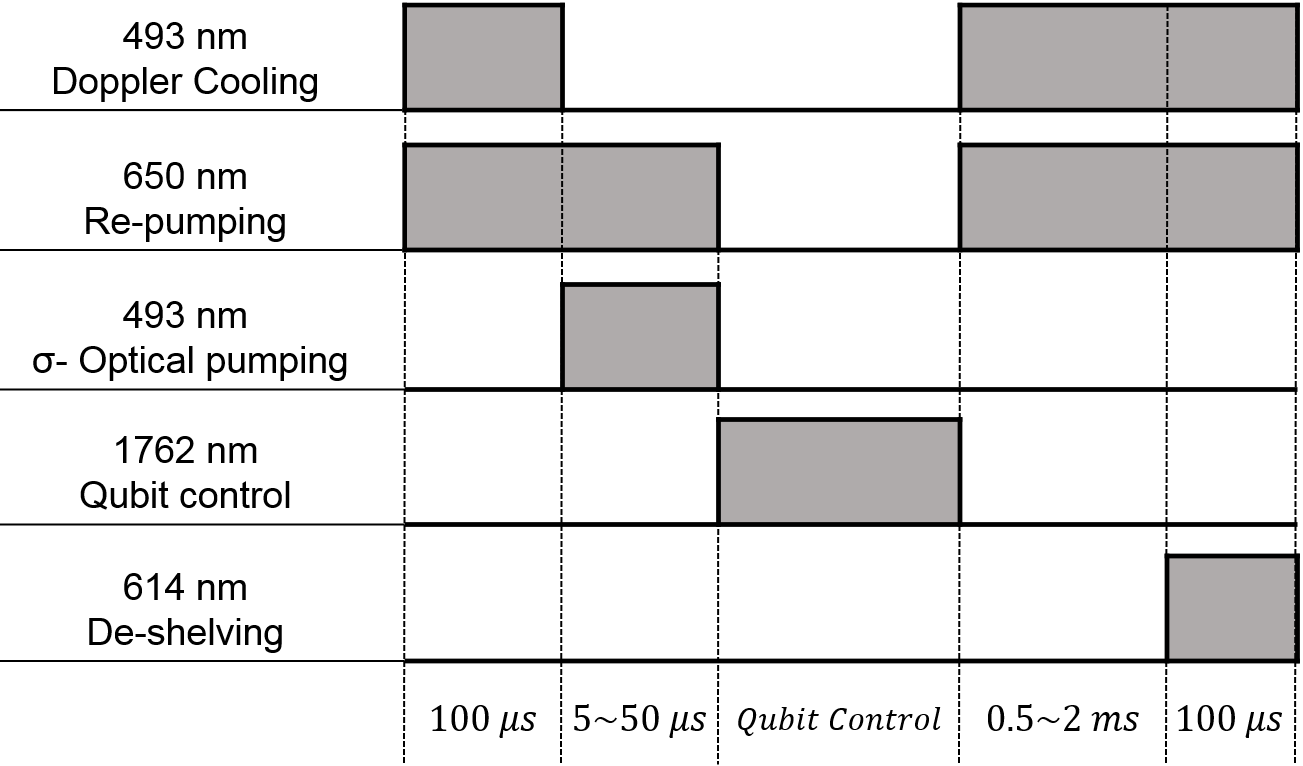}
\caption{Experimental sequence of lasers excitation pulses. The Ba$^+$ ion qubit is Doppler cooled and initialized by an optical pumping pulse as shown by the first three sequences. The $1762~$nm laser pulse is then used to transfer coherently the population from the initialized ground state to the excited qubit state. The probability of the transfer is then measured by detecting fluorescence light from the ion driven by the $493~$nm and $650~$nm laser shined in for about $1~$ms. At the end of the whole cycle, the excited state is de-shelved by a short pulse of on-resonant $614~$nm laser.}
\label{seq}
\end{figure}

The ultra low expansion (ULE) cavity is a horizontal design cavity from italics with a design stability of about one Hertz\cite{Gill,Millo,Bergquist}. The measured cavity linewidth is $1~$kHz which is within the specification of the manufacturer. The cavity lock electronics developed by us provides linewidth reduction of about $100$, therefore the estimated linewidth of the laser after locking to the ULE cavity is about few tens of Hz. However this has not been measured due to the non-availability of similar linewidth laser to perform heterodyne beat measurement, instead atomic spectroscopy on narrow transition has been performed which will be discussed in the following section.

The detection of a barium ion qubit state is performed by fluorescence measurement on the fast dipole transition at $493~$nm. The typical quantum efficiency of a  photomultiplier tube (PMT) is sensitive at this wavelength. Therefore discriminating qubit $\ket{0}$ or $\ket{1}$ state is not only done with a high efficiency but also in short time. In order to increase the collection efficiency of the fluorescence photons, a single aspheric lens which has a numerical aperture of $0.4$ mounted on a 3D-translation stage is placed inside the vacuum chamber.

All the laser pulses are controlled by a field programmable gate array (FPGA) based pulse generator setting the timing for the direct digital synthesizer (DDS) based radio-frequency generators. The control software is based on python and easily integrable to other systems. Further detail about the experimental software as well as hardware can be found in \cite{tarun,Deb}. In the following the single qubit gate performance has been characterized in more details. 

\section{Experimental Results}

As pointed out by the DiVincenzo's criteria \cite{DiVi}, a qubit is generally characterized by the following properties:

\begin{itemize}

\item[1] Initialization possibility: It should be possible to initialize the state of a qubit with high precision and repeatability so that repeated experiment can be performed on the same initial state. 
\item[2] Fast qubit operation: The qubit manipulation time needs to be faster than the lifetime of the qubit to avoid any significant decoherence of the state. 
\item[3] life time: The longer the qubit life-time is, the more is the number of error free quantum gates operations.
\item[4] High fidelity operations: The qubit manipulation needs to be performed with high fidelity such that error accumulation by the operation is negligible. 
\item[5] State discrimination: The final quantum state determination is inherently probabilistic in nature. Therefore many repetition of the same experiment needs to be performed before the probability of a state occupancy is finally determined. Hence to minimize the time of a single experiment, the discrimination between the qubit states needs to be high.

\end{itemize}

In the following, each of these characteristics for an optical qubit of a single barium ion in the telecommunication wavelength are discussed along with their present limitations.

\begin{figure}
\includegraphics[width=9cm]{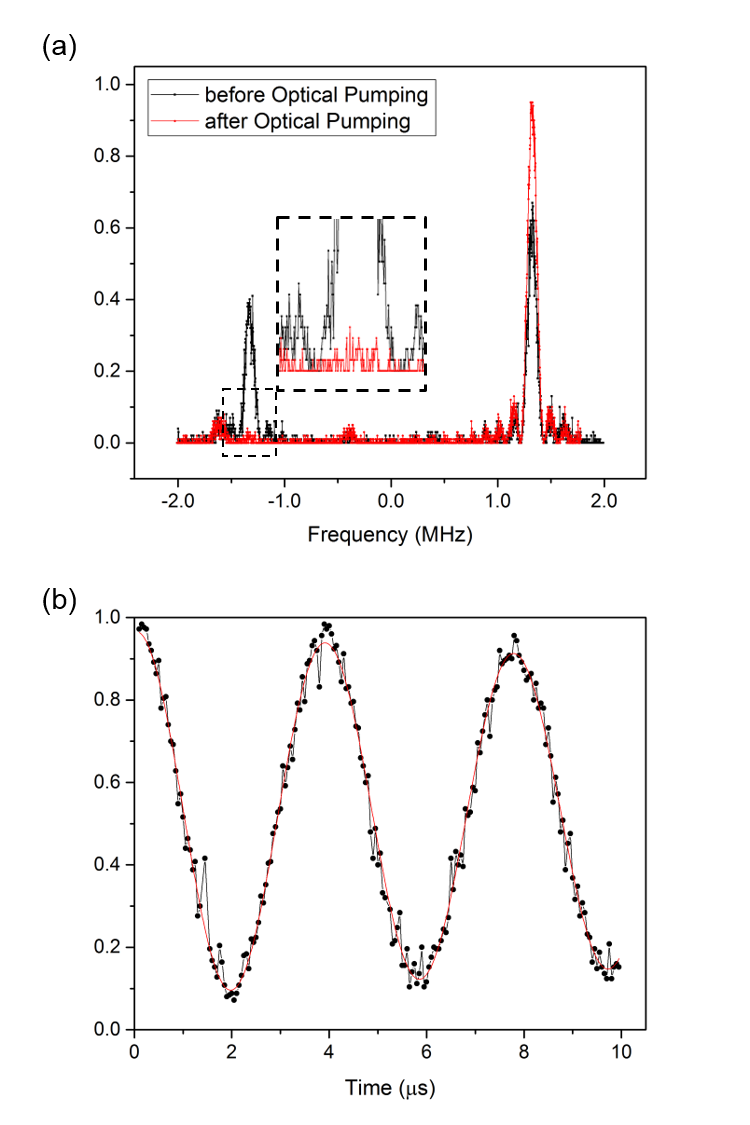}
\caption{(Color Online) (a) Spectra of the $S_{1/2}$ to $D_{5/2}$ quadrupole transition for $\Delta m = 0$ before and after optical pumping. The resonance peak from $\ket{6S_{1/2},m_{j}=+1/2}$ Zeeman level of ground state vanishes after optical pumping. The inset shows the remaining population at the unwanted dark ground state after optical pumping. This data shows $98.6~\%$ optical puming efficiency, limited by off-resonant scattering mostly due to impure polarization state of the optical pumping laser. (b) Resonance Rabi oscillation for the $\ket{S_{1/2},m_{j}=-1/2}~$ to $\ket{D_{5/2},m_{j}=-1/2}~$ transition. The fastest Rabi frequency obtained is $256.83~$kHz which is limited by the laser intensity noise.}\label{fig2}
\end{figure}

\subsection{Qubit initialization}

In Ba$^{+}$ ion optical qubit system, the 6$S_{1/2},m=-1/2$ state is defined as the $\ket{0}$ and the 5$D_{5/2},m=-1/2$ state is defined as $\ket{1}$. The experimental cycle as shown in fig.~\ref{seq}, in a typical single qubit operation starts with Doppler cooling and repumping of any population left in $\ket{1}$ state followed by optical pumping to $\ket{0}$ state. Once these two initialization steps are performed the rest of the cycle is used for rotating the qubit in the Bloch sphere and finally the measurement of the final state is performed. It is crucial to perform the optical pumping with high fidelity meaning ideally $100\%$ population in the state $\ket{0}$. In order to measure the error in our optical pumping step, the spectra of the dark Zeeman level $\ket{6S_{1/2},m_{j}=1/2}$ to $\ket{5D_{5/2},m_{j}=1/2}$ has been measured along with the spectrum of $\ket{6S_{1/2},m_{j}=-1/2}$ to $\ket{5D_{5/2},m_{j}=-1/2}$ qubit transition. This data is plotted in Fig. \ref{fig2} before and after the optical pumping step. As expected, once the optical pumping is successful, the dark Zeeman level $\ket{6S_{1/2},m_{j}=1/2}$ will be nearly empty and hence a ratio of the area under the two curves provide the error in preparing the initial state. The measured optical pumping efficiency in our case is $98.6~\%$. The error is mostly contributed by the spurious polarization left in the optical pumping beam as it is presently limited by the quality of the waveplate used.   
                                  
\subsection{Speed of operation}

The speed of quantum manipulation or single qubit gate operation time sets the limit for the speed of loading a memory qubit or single qubit phase gates. The speed in a single qubit gate is completely defined by the Rabi frequency of the carrier transition which is nothing but the speed at which a two level system is coherently driven by the excitation laser. Therefore the basic requirement is the intensity and linewidth of the driving laser. In our case, as mentioned earlier, the diode laser locked to the ULE cavity provides narrow linewith and hence the spectral density for the full power falling on the ion is significantly high. In addition, the aspheric lens allows tight focusing of the $1760~$nm light on to the ion even though its design wavelength is near $500~$nm. A measured Rabi oscillation on resonance is plotted in Fig.\ref{fig2}(b) which corresponds to $256.83~$kHz. As compared to other optical qubits in calcium or strontium it is similar in speed with the advantage that it can be used in conjunction with available optical telecom fiber networks. The figure shows that the contrast is lower than $100\%$ pointing to the fact that laser intensity or phase fluctuations are the dominant source of decoherence at these fast gate operation.

\subsection{Life time of the qubit}

\begin{figure}
\includegraphics[width=9.0cm]{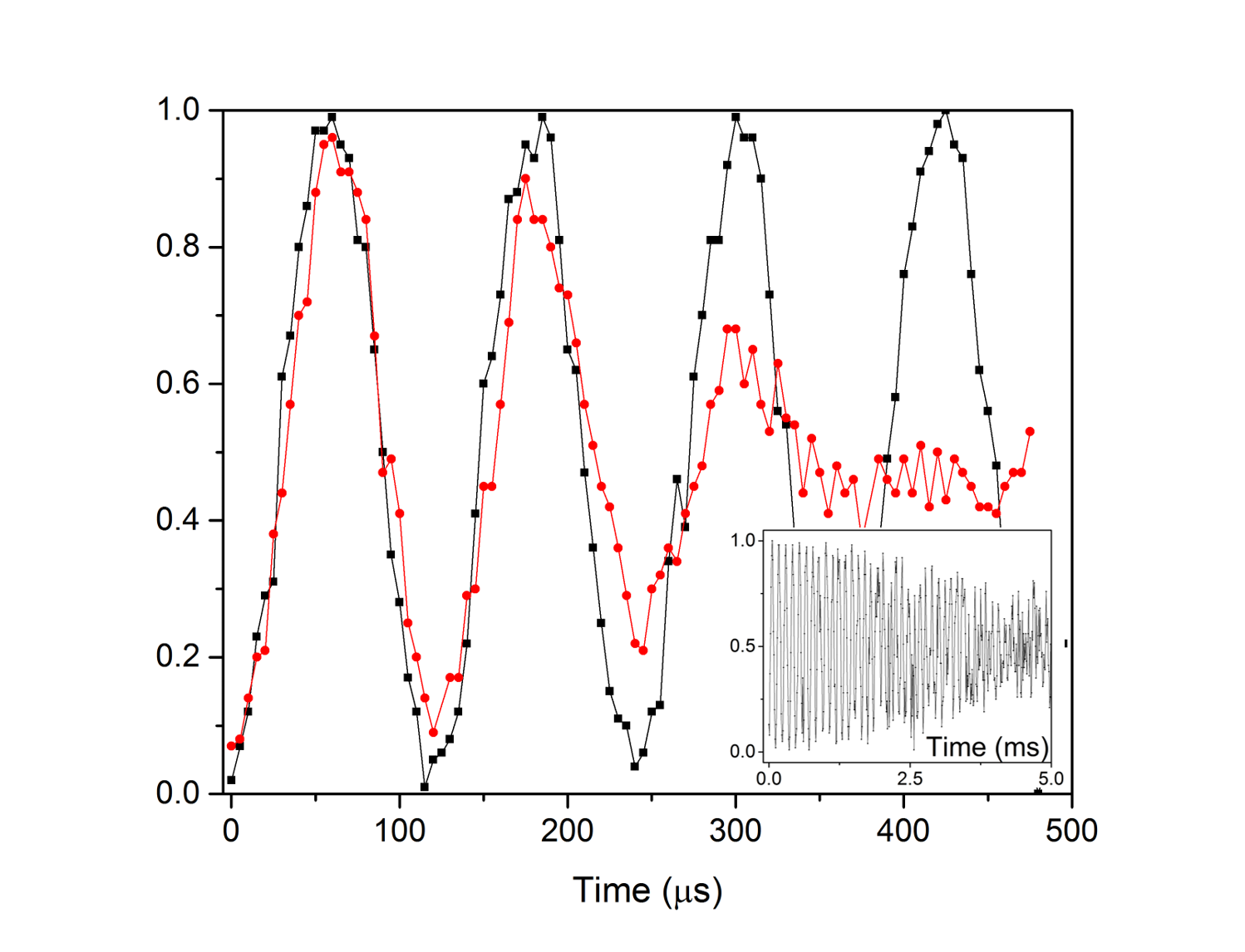}
\caption{(Color online) Ramsey measurement before (red) and after (black) $50$Hz line trigger. As observed from the decay of the contrast of the Ramsay fringes, $246(14)~\mu s$ coherence time (red) could be improved to over $3~$ms (black) after triggering the experimental sequence to a particular phase of the line frequency. (inset: Long time data of the Ramsey fringes after line triggering.)}
\label{fig3}
\end{figure}

The life time of a qubit means the useful time during which a qubit can be used for computation or storage. Therefore for scalability of any quantum computer or memory it is desired to have a long lifetime compared to gate operation time. The longest coherence time in a trapped ion qubit without any state preservation technique is in the order of $10~$s which is obtained with a ground state hyperfine qubits driven by Raman type two wavelength lasers \cite{long2005}. On the contrary lifetimes of optical qubits are limited by the upper state life-time but can be manipulated by a single wavelength laser. However in the Ba$^{+}$ ion qubit the decay time of the 5$D_{5/2}$ state is about $30~$s which is relatively long. To measure its coherence time, Ramsey-measurements is performed on the weakly sensitive $\Delta m=0$ transition between $\ket{6S_{1/2},m_{j}=-1/2}$ to $\ket{5D_{5/2},m_{j}=-1/2}$ levels as shown in Fig.\ref{fig3}. The observed coherence time is $246(14)~\mu$s which is mainly dominated by an ambient magnetic field fluctuations caused by the line frequency at $50~$Hz. We repeated the same experiment by synchronizing the experimental cycle to line frequency and we obtain coherence times over $3~$ms for $50\%$ loss in contrast. This as compared to the Rabi oscillation allows enough number of single qubit operation without significant loss in fideltity or contrast. There are two ways by which the qubit lifetime can be further improved, one is to passively shield the experimental vacuum chamber to protect from the environmental magnetic field noise\cite{Ruster} and the other is to apply successive spin echo or dynamic decoupling pulses \cite{Hahn1950,Lucas2011}.

\subsection{Fidelity}

\begin{figure}
\includegraphics[width=8.5cm]{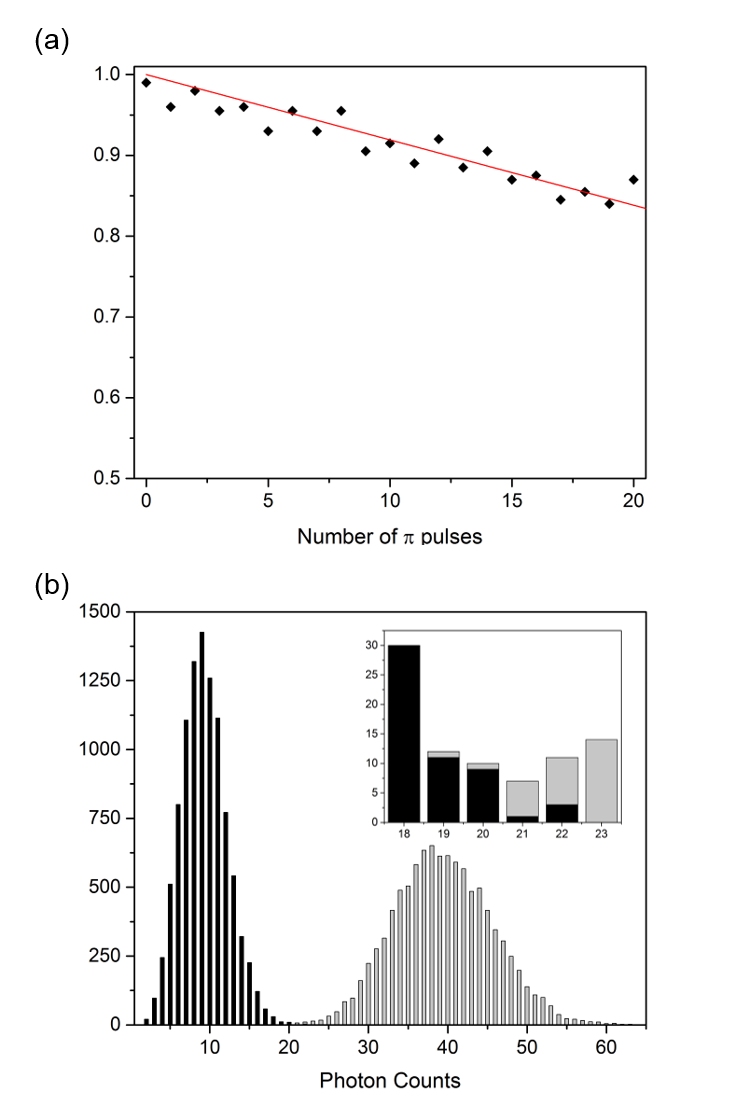}
\caption{ (a) Probability of the upper state occupancy after a series of $(2n+1)\pi$ pulses as a fucnting of $n$. The plotted dots are the measurement results and the solid line is a linear fitting which shows that the fidelity of a single $\pi$ pulse operation is $99.2~\%$. (b) The histograms of the photon counts as measured for two states of the qubit: the on state $\ket{1}~$ (Dark) and in the off state $\ket{0}~$ have medians at $40$ and $10$ counts respectively. The inset shows the overlap resion of the two histogram. In this region 40 events are mixed out of total 20000 runs, corresponding to $0.2~\%~$discrimination error.}\label{fidel}
\end{figure}

As a good and usable single quit, each rotation of the qubit needs to be performed with a high fidelity so as to avoid error propagation when the number of single qubit operations becomes large. Ideally, rotation in a Hilbert space is an unitary evolution, one expects it to be reversible, therefore the overlap between the initial state and the final state after $2\pi$ rotation is $100\%$. Any deviation is an error which needs to be accounted for. We obtained $99.2~\%$ fidelity by performing a series of $\pi$ pulses. The fidelity test result is shown at Fig.\ref{fidel}(a). It is likely that the external magnetic field fluctuations are the dominant source of fidelity loss in our system which is expected to be improved by passively shielding the experimental vacuum chamber from external magnetic field noise.

\subsection{State discrimination}

Once a single qubit operation is performed, the final state of the qubit needs to be determined. Since a measurement is inherently an occupation probability determination, repeated measurements on a projected state is required. Each time the qubit is prepared in the same initial state and the same operation is performed before the final state is projected on to $\ket{0}$ or $\ket{1}$. In an ion trap the optical qubit state is determined by fluorescence measurement and discrimination is done based on the observed photon counts. Therefore the error on the discrimination level determines the error on the state determination. In principle the threshold can be determined quite accurately if the {\it on} state fluorescence level which is dependent on the fluorescence collection time is set high. However longer measurement time leads to longer total operation time. Therefore, a high signal-to-noise ratio is necessary to keep the measurement time short. In our experiment the measurement time is $1~$ms with $0.2~\%$ discrimination error which is shown at Fig.\ref{fidel}(a). This error may be further improved by improving the signal-to-noise ratio which is presently optimized such that the total photon collection is high.      

\section{Conclusion}
Herein we have demonstrated an optical qubit on a quadrupole transition of a Ba$^{+}$ ion 6$S_{1/2}$ to 5$D_{5/2}$ transition operated by a stable and narrow linewidth diode based laser. The optical qubit fullfills the qualifications of an useful qubit characteristics, i.e. initialization possibility, fast qubit operation, long coherence time, high fidelity operation and well resolved state discrimination. The optical qubit can be extended to mulit-qubit system and requires only one addressing laser as compared to an hyperfine or Zeeman qubit addressed by Raman transition schemes. One could use microwave field but individual addressing in these long wavelength can only be performed in frequency domain with large magnetic field gradients\cite{Florian}. Therefore the Ba$^{+}$ ion optical qubit as demonstrated here can be combined with the existing telecommunication networks for quantum communications especially as nodes in networks.

\section{Acknowledgments}
This work is supported by Singapore Ministry of Education Academic Research Fund Tier 2 (Grant No. MOE2014-T2-2-119).

\section{Author's Contribution}
DY and DDM have conributed equally in this research.

\end{document}